\documentstyle[12pt,psfig]{article}
\begin{document}
\baselineskip 19pt

\begin{center}
     {\bf ULTRARELATIVISTIC POSITRONIUM PRODUCTION\\
	  IN COLLISIONS OF HIGH ENERGY ELECTRONS\\
		    AND LASER PHOTONS}\\[5mm]
     Martynenko A.P.\\
     Samara State University, Samara, Russia\\[3mm]
     Saleev V.A.\\
     Samara State University\\
     and Samara Municipal Nayanova University, Samara, Russia
\end{center}

\begin{abstract}
We consider the production of ultrarelativistic parapositronium
and orthopositronium atoms in collisions of high energy electrons
($E_e>0.5$ TeV) and laser photons. Our results demonstrate
the opportunity of intensive positronium beam formation with
Lorentz--factor $\gamma \sim 10^6$ using laser beam conversion
on high energy electrons at the future $e^+e^-$ Linear
Accelerators.
\end{abstract}

\section*{Introduction}

Formation of ultrarelativistic positronium beams is of immediate
interest as a test of main principles of quantum electrodynamics
(QED) ~\cite{1}
and theory of relativistic bound states  ~\cite{2}.
In particular, the formation of ultrarelativistic parapositronium
($^1S_0$ -- state) and
P-wave positronium beams with very large Lorentz--factors $\gamma$
gives a chance to measure with high accuracy their decay widths. It is
needed because of theoretical predictions at present time are more
precise than experimental data ~\cite{3}.
At very large Lorentz--factor  $\gamma \sim 10^6$
the positritronium formation length is about several sm
for ground states ($n=1$, $^3S_1$ or $^1S_0$) ~\cite{4}
and it is realty to check directly the exponential law for
the $e^+e^-$ bound state formation in vacuum and matter.

At present time it is discussed different approach for obtaining
relativistic positronium beams. First of all, this is positronium
production in $\pi^0$--meson decays at Proton Accelerators,
which have been tested experimentally some years ago ~\cite{5}.

Relativistic positronium beams may be formed in high energy photon
~\cite{6} or electron ~\cite{7} interaction with  matter.
Recently it was predicted positronium production rate in collisions of
heavy ions ~\cite{8}.

Here we consider the formation of ultrarelativistic
positronium beams ($\gamma \simeq 10^6$ )
via laser photon conversion on high energy electrons.
This method may be realized at future Linear
$e^+e^-$ Accelerators at the energy range $E_e \approx 1$ TeV.

The idea of our calculation is the same one which has been discussed
previously ~\cite{9} in the case of realizing of
$\gamma e$ -- and $\gamma \gamma$ -- collisions on the basis of Linear
$e^+e^-$ Collider.
The high density laser beam (approximately $10^{20}$ photons per
impulse with the energy $\omega=1$ eV)
backscatteres on high energy electron beam of Linear Accelerator and
converts into high energy photon beam with probability equal to unity.
At the laser power about $10$ J and frequency of laser impulse
divisible to Linear Collider frequency
( $f\sim 10 \div 100\  \mbox s^{-1}$ ) it will be so many laser photons
that almost high energy electron transfers energy to laser photon.

It is obviously that at conditions discussed above, the ratio
(\ref{eq:al1}):
\begin{equation}
{\cal K}=\frac{\sigma(\gamma+e \rightarrow Ps+e)}
{\sigma(\gamma+e \rightarrow \gamma+e)},\label{eq:al1}
\end{equation}
fixed the number of positronia per one electron in the initial beam.
There will hard energy spectrum of positronia flying in the narrow
cone near the initial electron beam direction, as well as in the case of
converted photons.

\section{The process $\gamma+e \rightarrow Ps+e$}

In the nonrelativistic approximation, positronium is considered as a
$e^+e^-$ -- system with the fixed mass $M=2m$, zero binding energy and zero
relative momentum. To project the pair of free electron and positron
on the $^3{S}_1$ or $^1{S}_0$ bound states we have used the next
projection operators ~\cite{10}:
\begin{eqnarray}
\hat P(^3{S}_1)=\frac{\Psi(0)}{\sqrt{2m}}\hat\varepsilon(\hat p/2+m),\\
\hat P(^1{S}_0)=\frac{\Psi(0)}{\sqrt{2m}}\gamma_5(\hat p/2+m),
\end{eqnarray}
where $\hat\varepsilon=\varepsilon^\mu(p)\gamma_\mu,\
\varepsilon^\mu(p)$ is the orthopositronium polarization four--vector,
$p$  is the positronium four--momentum,
$\Psi(0)=\sqrt{m^3\alpha^3/8\pi}$  is the nonrelativistic positronium
ground state wave function at the origin,  $m$ is the electron mass.

In the lowest order in $\alpha=e^2/4\pi$ positronium production in the
process $\gamma+e\rightarrow Ps+e$ is described by the Feynman
diagrams in Fig 1. The nonzero contribution comes from diagrams
1-6 for parapositronium production and from diagrams 1-4, 7, 8 for
orthopositronium production:
\begin{eqnarray}
M_1&=&e^3\bar U(q_2)\gamma^{\mu}\hat P\gamma_{\mu}(\hat q_1+\hat k +m)
     \hat\varepsilon (k)U(q_1)/\nonumber\\
&&((q_1-k)^2-m^2)(p/2-q_2)^2 \\
M_2&=&e^3\bar U(q_2)\gamma^{\mu}\hat	P\hat	\varepsilon   (k)(\hat
p/2-\hat k+m)\gamma_{\mu}U(q_1)/\nonumber\\
&&((p/2-k)^2-m^2)(q_2+p/2)^2\\
M_3&=&e^3\bar U(q_2)\hat \varepsilon (k)(\hat q_2-\hat k+m)\gamma^{\mu}
     \hat P\gamma_{\mu}U(q_1)/\nonumber\\
&&((q_2-k)^2-m^2)(q_1-p/2)^2\\
M_4&=&e^3\bar U(q_2)\gamma^{\mu}(-\hat p/2+\hat k+m)\hat\varepsilon (k)
     \hat P\gamma_{\mu}U(q_1)/\nonumber\\
&&((p/2-k)^2-m^2)(q_1-p/2)^2\\
M_5&=&e^3\bar U(q_2)\gamma^{\mu} U(q_1)
     \mbox{Tr}[\hat P\hat\varepsilon  (k)(-\hat   k+\hat   p/2+m)
\gamma_{\mu}]/\nonumber\\
     &&(q_1-q_2)^2((p/2-k)^2-m^2)\\
M_6&=&e^3\bar U(q_2)\gamma^{\mu}U(q_1)\mbox{Tr}[\hat P\gamma_{\mu}
     (-\hat p/2+\hat k+m)\hat \varepsilon (k)]/\nonumber\\
    && (q_1-q_2)^2((p/2-k)^2-m^2)\\
M_7&=&e^3\bar U(q_2)\gamma^{\mu}(\hat q_1+\hat k+m)\hat\varepsilon (k)
     U(q_1)\mbox{Tr}[\gamma_{\mu}\hat P]/\nonumber\\
    &&p^2((q_1+k)^2-m^2)\\
M_8&=&e^3\bar U(q_2)\hat\varepsilon (k)(-\hat p+\hat q_1+m)U(q_1)
     \mbox{Tr}[\gamma_{\mu}\hat P]/\nonumber\\
     &&p^2((p-q_1)^2-m^2)
\end{eqnarray}

Let us define the Mandelstam variables for the process $\gamma+e \rightarrow
Ps+e$ in the case of $E_e\gg m\gg\omega\sim 1$ eV:
\begin{eqnarray}
s&=&(q_1+k)^2 \simeq m^2+4\omega E_e,\\
t&=&(k-p)^2\simeq 4m^2-2E_e\omega(1+\cos\theta),\\
u&=&(q_1-p)^2\simeq 5m^2-2E_eE(1-\cos\theta),
\end{eqnarray}
where $k=(\omega,0,0,-\omega)$ is laser photon four--momentum,
$q_1=(E_e,0,0,E_e)$ is the initial electron four--momentum,
$q_2$ is the scattered electron four--momentum,
$p=(E,0,E\sin\theta,E\cos\theta)$ is positronium the four--momentum,
$E$ is the energy of positronium,
$E_e$ is the energy of initial electron,
$\theta$ is scattering angle of positronium,
\begin{equation}
\cos\theta\simeq 1-\frac{2\omega E_e+2m^2}{EE_e}.
\end{equation}
The differential cross section for the process $\gamma+e \rightarrow
Ps+e$ as a function of
$y=E/E_e\simeq (4m^2-t)/(s-m^2)$ is expressed in terms of
$\overline{ |M|^2}$ as follows:
\begin{equation}
\frac{\mbox{d}\sigma}{\mbox{d}y}(\gamma+ e\rightarrow Ps+e)=
\frac{\overline{|M|^2}}{16\pi(s-m^2)}\label{eq:al7}
\end{equation}
The total cross section $\sigma(\gamma+ e\rightarrow Ps+e)$ is obtained from
(~\ref{eq:al7}) integrating with respect to $y$ in the limits:
\begin{eqnarray}
y_{max\atop min}=\frac{1}{s-m^2}\Biggl[2m^2&+&\frac{(s+m^2)(s-3m^2)}{2s}
\nonumber\\&&\pm\frac{(s-m^2)}{2s}\sqrt{(s-9m^2)(s-m^2)}\Biggr]
\end{eqnarray}

\section{Results of calculations}
First of all, to note that minimal value of invariant $s$
in the process $\gamma+e \rightarrow Ps+e$ is $s_{min}=9m^2\approx
2.35 \mbox{MeV}^2 $
and corresponding threshold electron energy is equal to
\begin{equation}
E_{e,min}=\frac{s_{min}-m^2}{4\omega}=\frac{2m^2}{\omega}.
\end{equation}
At the energy of laser photons $\omega=1$ eV we obtain from (18)
$E_{e,min}=522$ GeV. On the other hand, at $E_e=1$ TeV (the energy
range of future Linear Colliders)
$s$ is equal to 4.2 MeV$^2$. Figure 2 shows the calculated total cross
sections
for orthopositronium (curve 1) and parapositronium (curve 2) production
as a function of $s$ at $\omega$=1 eV. The orthopositronium cross section
has a maximum at
$s=3.2$ MeV$^2$ (or $E_e=735$ GeV), where one has
$\sigma(\gamma+ e\rightarrow {^3S_1}+e)\approx 237\mbox{ pb}$ and
$\sigma(\gamma+ e\rightarrow {^1S_0}+e)\approx 325\mbox{ pb}$.

The orthopositronium production cross section decreases at large $s$
opposite to parapositronium production cross section which increases
logarithmically. This behavior comes from diagrams 4 and 5 in Fig 1.
At the electron energy $E_e=1$ TeV and $\omega =1$ eV
($s\approx 4.2$ MeV$^2$) we have found that $\sigma(\gamma+ e
\rightarrow ^3{S}_1+e)
\approx 218$ pb and $\sigma(\gamma +e\rightarrow ^1{S}_0+e)
\approx 494$ pb.

The above mentioned values of cross sections correspond to production of
positronia in ground states with $n=1$.
The summation over $n$ enhances the obtained results by a factor of
$\zeta (3)=\sum_{n=1}^{\infty}1/n^3\simeq
1.202$.

Table 2 shows the values of ratio ${\cal K}$ (1) at different $s$.
In such a manner, at $E_e=1$ TeV and at number of electrons per impulse
$n=2\cdot10^{11}$ (the project of $e^+e^-$ Linear Collider VLEPP) it will be
produced $\sim 1200$ parapositronium and $\sim 500$ orthopositronium per
impulse. At the accelerator frequency $f=100\ s^{-1}$ it gives
$1.2\cdot 10^5$ and $0.5\cdot 10^5$ positronium atoms per second.
This result exceeds the number of positronium atoms which may be
obtained in the
recombination process $e^++e^-\rightarrow Ps+\gamma$ at
$e^+e^-$ Storage Rings ~\cite{11} at the same parameters of accelerators.
The rough estimation on
$P$--wave positronium production rate in the process
$\gamma+e \rightarrow Ps+\gamma$ is approximately
$10^{-4}$ from the number of $S$--wave positronium production rate, i.e.
$\sim 100$ atoms per second. It seems, that so large $P$--wave positronium
production rate will enough for precise test of their decay widths.

The positronium spectra $\mbox d N/\mbox d y (y,s)$ normalized to unity at
$s=3.2$ MeV$^2$ are shown in Fig.1. The positronium atoms are produced
in the kinematic range
$0.375\leq y\leq 0.869$ with the average values $<y>\approx 0.54$ for
parapositroniums and  $<y>\approx 0.63$ for orthopositroniums and they have
ultrarelativistic Lorenz--factors
$0.27\cdot 10^6\leq \gamma\leq 0.64\cdot 10^6$.

In conclusion we note that, the obtained via laser photon conversion
on high energy electrons, ultrarelativistic positronium beam
will be pure, from the point
of view of hadronic background. It follows from the trivial kinematic
fact that threshold energy of pair $\pi$--meson production is
 $E_{min}\approx 2\cdot 10^4$ TeV at $\omega\approx 1$ eV.

We are grateful to  E.A.~Kuraev,
L.L.~Nemenov, I.N.~Meshkov, V.G.~Serbo, R.N.~Faustov and I.B.~Khriplovich
for useful discussion of positronium atom physics. This work was done under
the financial support of the Russian Foundation for Basic Researches
(Grants 98-02-16185 and
98-15-96040) and Programme ''Universyties of Russia'' (Grant 2759).


\newpage

\section*{\bf Table 1}

$${\cal K}=\sigma(\gamma+ e \rightarrow Ps+e)/
\sigma(\gamma +e \rightarrow\gamma +e)$$
\begin{center}
\begin{tabular}{|c|c|c|}\hline
$s$,MeV$^2$ & $^3{S}_1$ & $^1{S}_0$ \\ \hline
$3.2$ & $2.35\cdot 10^{-9}$ & $3.23\cdot 10^{-9}$ \\ \hline
$4.2$ & $2.61\cdot 10^{-9}$ & $5.88\cdot 10^{-9}$ \\ \hline
\end{tabular}
\end{center}
\vspace{5mm}

\section*{Figure captions}

\begin{enumerate}
\item Diagrams used for the process $\gamma +e\rightarrow Ps+e$.
\item Cross section for the process $\sigma(\gamma +e\rightarrow Ps+e)$
as a function of $s$ at laser photon energy $\omega=1$ eV.
Here $Ps$ is the orthopositronium (curve 1) and parapositronium (curve 2).
\item The positronium spectra $\mbox{d}N/\mbox{d}y$  normalized to unity
in the process $\gamma +e\rightarrow Ps+e$, where $y=E/E_e$. Curve 1
is the orthopositronium spectrum, curve 2 is the parapositronium spectrum.
\end{enumerate}
\newpage

\begin{figure}[p]
\unitlength 1mm
\psfig{figure=pic12.ps,height=20cm,width=10cm,%
	bbllx=2cm,bblly=5cm,bburx=12cm,bbury=25cm}%
\vspace*{1.0cm}
\end{figure}

\newpage

\begin{figure}[p]
\unitlength 1mm
\psfig{figure=pic34.ps,height=20cm,width=10cm,%
	bbllx=2cm,bblly=5cm,bburx=12cm,bbury=25cm}%
\vspace*{1.0cm}
\end{figure}

\newpage
\begin{figure}[p]
\unitlength 1mm
\psfig{figure=pic56.ps,height=20cm,width=10cm,%
	bbllx=2cm,bblly=5cm,bburx=12cm,bbury=25cm}%
\vspace*{1.0cm}
\end{figure}

\newpage

\begin{figure}[p]
\unitlength 1mm
\psfig{figure=pic78.ps,height=20cm,width=10cm,%
	bbllx=2cm,bblly=5cm,bburx=12cm,bbury=25cm}%
\vspace*{1.0cm}
\end{figure}

\newpage

\begin{figure}[p]
\unitlength 1mm
\psfig{figure=ega.ps,height=20cm,width=10cm,%
	bbllx=2cm,bblly=5cm,bburx=12cm,bbury=25cm}%
\vspace*{1.0cm}
\end{figure}

\newpage

\begin{figure}[p]
\unitlength 1mm
\psfig{figure=dy.ps,height=20cm,width=10cm,%
	bbllx=2cm,bblly=5cm,bburx=12cm,bbury=25cm}%
\vspace*{1.0cm}
\end{figure}

\end{document}